\documentstyle{article}
\begin{document}

\baselineskip=7mm

\noindent
SAGA-HE-88-95

\noindent
July 1995

\bigskip

\bigskip

\noindent
Incompressibility of nuclear matter, and Coulomb and volume-symmetry

\noindent
coefficients of nucleus incompressibility
in the relativistic mean field theory

\centerline{\bf{H. Kouno, T. Mitsumori, N. Noda, K. Koide, A. Hasegawa}}

\centerline{Department of Physics, Saga University, Saga 840, Japan}

\centerline{\bf{and}}

\centerline{\bf{ M. Nakano}}

\centerline{University of Occupational and Environmental Health, Kitakyushu
807, Japan}

\bigskip

\bigskip

\centerline{(PACS numbers: 21.65.+f, 21.30.+y )}

\centerline{\bf ABSTRACT}

\noindent
The volume coefficient $K$(=incompressibility of the nuclear matter), the
Coulomb coefficient $K_c$, and the volume-symmetry coefficient $K_{vs}$ of the
nucleus incompressibility are studied in the framework of the relativistic mean
field theory, with aid of the scaling model.
It is found that $K= 300\pm 50$MeV is necessary to account for the empirical
values of $K_v$, $K_c$, and $K_{vs}$, simultaneously. The result is independent
on the detail descriptions of the potential of the $\sigma$-meson
self-interaction and is almost independent of the strength of the
$\omega$-meson self-interaction.

\vfill\eject

One way to determine the incompressibility $K$ of nuclear matter from the giant
monopole resonance (GMR) data is using the leptodermous expansion[1] of nucleus
incompressibility $K(A,Z)$ as follows.
$$  K(A,Z)=K+K_{sf}A^{-1/3}+K_{vs}I^2+K_cZ^2A^{-4/3}+\cdot \cdot
\cdot~~~~~;~~~I=1-2Z/A, \eqno{(1)} $$
where the coefficients, $K$, $K_{sf}$, $K_{vs}$ and $K_c$ are volume
coefficient (incompressibility of nuclear matter), surface term coefficient,
volume-symmetry coefficient and Coulomb coefficient, respectively.
We have omitted higher terms in eq. (1).
Although there is uncertainty in the determination of these coefficients by
using the present data, Pearson [2] pointed out that there is a strong
correlation between $K$ and $K_c$.
(See table I.)
Similar observations are done by Shlomo and Youngblood [3].


\centerline{$\underline{~~~~~~~}$}

\centerline{Table I}

\centerline{$\underline{~~~~~~~}$}


According to this context, Rudaz et al. [4] studied the relation between
incompressibility and the skewness coefficient by using the generalized version
of the relativistic Hartree approximation [5].
The compressional and the surface properties are studied by Von-Eiff et al.
[6-8] in the framework of the relativistic mean field approximation of the
$\sigma$-$\omega$-$\rho$ model with the nonlinear $\sigma$ terms.
They found that low incompressibility ($K\approx 200$MeV) and a large effective
nucleon mass $M^*$ at the normal density ($0.70\leq M^*/M\leq 0.75$) are
favorable for the nuclear surface properties [8].
On the other hand, using the same model, Bodmer and Price [9] found that the
experimental spin-orbit splitting in light nuclei supports $M^*\approx 0.60M$.
The result of the generator coordinate calculations for breathing-mode GMR by
Stoitsov, Ring and Sharma [10] suggests $K\approx 300$MeV.

In previous papers[11,12], we have studied the relation between $K$ and $K_c$
in detail, using the relativistic mean field theory with the nonlinear $\sigma$
terms [13] and the one with the nonlinear $\sigma$ and $\omega$ terms [14].
We found that, under the assumption of the scaling model [1], $K=300\pm 50$MeV
is favorable to account for $K$, $K_c$ and $K_{vs}$, simultaneously.
It seems that this conclusion is not drastically changed in the use of the
relativistic mean field theory and the scaling model.
In this paper, we examine the conclusion in more general way, in which the
result does not depend on the detail descriptions of the $\sigma$-meson
self-interaction.
The reason why we restrict our discussions to $K$, $K_c$ and $K_{vs}$ is that
the general discussions, which are independent of the detail of the model (
e.g., types of the interactions, values of the parameters in the Lagrangian,
etc.) are possible to a considerable extent, since, as is shown below, these
quantities are almost analytically estimated by using the result for the
nuclear matter, if we assume the scaling model [1].

We use the relativistic mean field theory based on the $\sigma$-$\omega$-$\rho$
model with the nonlinear $\sigma$ terms.
The Lagrangian density consists of four fields, the nucleon $\psi$, the scalar
$\sigma$-meson $\phi$, the vector $\omega$-meson $V_\mu$, and the
vector-isovector $\rho$ meson ${\bf b}_\mu$, i.e.,
$$ L_{N\sigma\omega\rho} =\bar{\psi}(i\gamma_\mu\partial^\mu -M)\psi$$
$$
+{1\over{2}}\partial_\mu\phi\partial^\mu\phi -{1\over{2}}m_s^2\phi^2
-{1\over{4}}F_{\mu\nu} F^{\mu\nu} +{1\over{2}}m_v^2V_\mu V^\mu
-{1\over{4}}{\bf{B}}_{\mu\nu}\cdot{\bf{B}}^{\mu\nu}+{1\over{2}}m_\rho^2{\bf{b}}_\mu\cdot{\bf{b}}^\mu$$
$$ +g_s\bar{\psi}\psi\phi-g_v\bar{\psi}\gamma_\mu \psi V^\mu
-g_\rho\bar{\psi}\gamma_\mu{\tau\over{2}}\cdot{\bf {b}}^\mu\psi-U(\phi )~~~;$$
$$F_{\mu\nu}=\partial_\mu V_\nu -\partial_\nu
V_\mu,~~~~~{\bf{B}}_{\mu\nu}=\partial_\mu {\bf{b}}_\nu -\partial_\nu
{\bf{b}}_\mu-g_\rho {\bf b}_\mu\times {\bf b}_\nu , \eqno{(2)} $$
where $m_s$, $m_v$, $m_\rho$, $g_s$, $g_v$ and $g_\rho$ are $\sigma$-meson
mass, $\omega$-meson mass, $\rho$-meson mass, $\sigma$-nucleon coupling,
$\omega$-nucleon coupling, and $\rho$-nucleon coupling, respectively.
The $U(\phi )$ is a nonlinear self-interaction potential of $\sigma$ meson
field $\phi$.
For example, in ref. [11], we have used the quartic-cubic terms of $\phi$ as in
ref. [13], i.e.,
$$
U(\phi ) ={{1}\over{3}}b\phi^3+{{1}\over{4}}c\phi^4, \eqno{(3)} $$
where $b$ and $c$ are the constant parameters which are determined
phenomenologically.
However, in this paper, we do not give an explicit expression of $U(\phi)$ and
discuss the problem in more general way, without any assumption of $U(\phi )$.

In the scaling model [1], $K$, $K_c$ and $K_{vs}$ in eq. (1), are given by
$$  K=9\rho_0^2{{\partial^2E_b}\over{\partial \rho^2}}\vert_{\rho=\rho_0},
\eqno{(4)} $$
$$  K_c=-{{3q_{el}^2}\over{5R_0}}\biggl( {9K'\over{K}}+8 \biggr), \eqno{(5)} $$
$$  K_{vs}=K_{sym}-L\biggl( 9{K'\over{K}}+6 \biggr), \eqno{(6)} $$
where $\rho$, $\rho_0$, $E_b$ and $q_{el}$ are the baryon density, the normal
baryon density, the binding energy per nucleon, and the electric charge of
proton, respectively, and $R_0=[3/(4\pi\rho_0)]^{1/3}$,
$$ K'=3\rho_0^3{{d^3 E_b}\over{d \rho^3}}\vert_{\rho=\rho_0}, \eqno{(7)} $$
$$  L=3\rho_{0}{{d J}\over{d \rho}}\vert_{\rho =\rho_{0}},
{}~~~~K_{sym}=9\rho_{0}^2{{d^2 J}\over{d \rho^2}}\vert_{\rho =\rho_{0}}:
{}~~~~J={1\over{2}}\rho^2{{\partial^2E_b}\over{\partial\rho_3^2}}\vert_{\rho_3=0}.  \eqno{(8)}  $$
The quantity such as $K'$ is sometimes called "skewness".

In the mean field theory with the Lagrangian (2), $L$ and $K_{sym}$ are given
by [7]
$$  L={{3\rho}\over{8M^2}}C_\rho^2+{\rho\over{2}}\biggl(
{{2k_Fk^{\prime}_F}\over{E_F^*}}-{{k_F^2E_F^{*\prime}}\over{E_F^{*2}}}\biggr)~~~~~~(\rho =\rho_0), \eqno{(9)} $$
and
$$  K_{sym}={3\over{2}}\rho^2\biggl(
{{2k_F^{\prime2}}\over{E_F^*}}+{{2k_Fk_F^{\prime\prime}}\over{E_F^*}}-{{4k_Fk_F^{\prime}E_F^{*\prime}}\over{E_F^{*2}}}+{{2k_F^2E_F^{*\prime 2}}\over{E_F^{*3}}}-{{k_F^2E_F^{*\prime\prime}}\over{E_F^{*2}}} \biggr)~~~~~(\rho =\rho_0),    \eqno{(10)} $$
where $k_F(=[3\pi^2\rho /2]^{1/3})$ is the Fermi momentum, and $C_\rho =g_\rho
M/m_\rho$,
$$
k'_F={{dk_F}\over{d\rho}}={k_F\over{3\rho}},~~~~~k_F^{\prime\prime}={{d^2k_F}\over{d\rho^2}}={{-2k_F}\over{9\rho^2}}, \eqno{(11)} $$
$$    E_F^*=\sqrt{k_F^2+M^{*2}}, ~~~~~E_F^{*\prime}={{dE_F^*}\over{d\rho}},
{}~~~~~E_F^{*\prime\prime}={{d^2E_F^*}\over{d\rho^2}}.   \eqno{(12)} $$
$M^*$ in eq. (12) is the effective nucleon mass.
Furthermore, at $\rho =\rho_0$, $E_F^{*\prime}$ and $E_F^{*\prime\prime}$ are
related to $K$ and $K'$ in the following relations, respectively [7,11].
$$
E_F^{*\prime}={{K}\over{9\rho_0}}-{{C_v^2}\over{M^2}};~~~~~C_v={{g_vM}\over{m_v}}. \eqno{(13)}   $$
$$    E_F^{*\prime\prime} ={{K+K'}\over{3\rho_0^2}}. \eqno{(14)} $$
At $\rho =\rho_0$, $C_v$ and $C_\rho$ are also related to $M^*$ as follows.
$$ C_v^2={{M^2}\over{\rho_0}}\biggl( M-a_1-\sqrt{k_F^2+M^{*2}}\biggr) ,
\eqno{(15)} $$
[13] and
$$  C_\rho^2 ={{8M^2}\over{\rho_0}}\biggl(
a_4-{{k_{F}^2}\over{6E_{F}^*}}\biggr), \eqno{(16)} $$
[15] where $a_1$ and $a_4$ are the binding energy and the symmetry energy at
$\rho =\rho_0$, respectively.
We remark that eqs. (5), (6), and (9)$\sim$(16) have no explicit dependence on
$U(\phi )$.

{}From eq. (5), $K'$ are determined, if $\rho_0$, $K$ and $K_c$ are given.
Therefore, from the eqs. (5), (6), and (9)$\sim$(16), it is seen that $K_{vs}$
is determined, if $\rho_0$, $M$, $a_1$, $a_4$, $K$, $K_c$, and $M^*$ are given,
without giving the detail descriptions for $U(\phi )$.
Using these equations, we calculate $K_{vs}$.
In the calculations, we put $\rho_0=0.16$fm$^{-3}$, $M=939$MeV, $a_1=15.75$MeV
and $a_4=30.0$MeV.
For $K$ and $K_c$, we use the values in table I.
We assume that $M^*=0.5M\sim 0.94M$, the phenomenologically acceptable values.
(We remark the upper bound for $M^*$ is gotten, if we put $C_v=0$ in eq. (15).
)
In fig. 1 (solid lines), we show $K_{vs}$ as a function of $M^*$ for two sets
of $K$ and $K_c$ in table I.
In the figures, $K_{vs}$ decreases as $M^*$ increases.
This is understood as follows.
As $M^*$ increases, $L$ and $K_{sym}$ decreases, because of large $E_F^*$ in
the denominators in eqs. (9) and (10).
Small $K_{sym}$ makes $K_{vs}$ smaller (more negative) and, on the contrary,
small $L$ makes $K_{vs}$ larger (less negative), since $(9K'/K+6)>0$ in eq.
(6).
In the cases of fig. 1, the effect of the small $K_{sym}$ overcomes that of the
small $L$, therefore, $K_{vs}$ decreases.
In the case of $K=300$MeV and $K_c=-3.990$MeV, $K_{vs}=-289\sim -98$MeV.
These values are in good agreement with the corresponding empirical values in
table I.
We also remark that the uncertainty of $K_{vs}$ in changing $M^*$ is comparable
to the empirical error bar of $K_{vs}$ in table I.
In the case of $K=250$MeV and $K_c=-0.7065$MeV, $K_{vs}=50\sim 410$MeV.
These values are somewhat larger than the empirical ones.
In table II(a), we summarize the range of the calculated $K_{vs}$ for each set
of $K$ and $K_c$ in table I.
Comparing the table I and table II(a), we see that $K=300\pm 50$MeV is
necessary to account for $K$, $K_c$ and $K_{vs}$, simultaneously.


\centerline{$\underline{~~~~~~~~~~~~~~~~~~~~~}$}

\centerline{Fig. 1(a),(b), Table II(a),(b)}

\centerline{$\underline{~~~~~~~~~~~~~~~~~~~~~}$}


We remark the following three points.

(1) The results are independent of the form of $U(\phi )$, since eqs. (5), (6),
and (9)$\sim$(16) are required for any type of $U(\phi )$.

(2) The question, whether there are coupling parameters, which reproduce the
set of $K$ and $K_c$ in table I, or not, is still open, and the answer for the
question depends on the detail descriptions of $U(\phi )$.
For example, if we use the quartic-cubic potential (3),
we could not find the coupling parameters (i.e., $g_s$, $g_v$, $b$,  and $c$ ),
which reproduce $K=200$MeV and $K_c=2.577$MeV [11].
(We remark that $K$ and $K_c$ are independent on $g_\rho$ in the mean field
theory. )
Also, using eq. (3), we get the parameter set for $K=300$MeV and
$K_c=-3.990$MeV, only in the case of $M^*=0.83M$ [11].
In those cases, parameter set is uniquely determined or could not be found,
since the number of the parameters is not larger than the number of the inputs,
i.e., $K$, $K_c$, and two conditions for the saturation.
If the higher terms of $\phi$ are added to (3), the wider range of $M^*$ may be
available.
However, $K=300\pm 50$MeV is $necessary$ to reproduce the empirical values of
$K$, $K_c$, and $K_{vs}$ simultaneously, for any type of $U(\phi )$.

(3) The results do not have a strong-dependence on $\rho_0$, $a_1$ and $a_4$,
since the calculated $K_{vs}$ is much more sensitive to the ratio $K'/K$ than
to those quantities.

Next we add the following attractive term of vector self-interaction (VSI)
[14,12] to the Lagrangian (2).
$$ L_{VSI}={{g_v^2Y^2}\over{4}}m_v^2(V_\mu V^\mu)^2, \eqno{(17)} $$
, where $Y$ is a positive constant which determine the strength of VSI.
Although, according to this modification, the formalism for calculating
$K_{vs}$, which is described above, is slightly modified, it is still possible
to study  $M^*$-$K_{vs}$ relation without giving the detail descriptions of
$U(\phi )$, as in the case of no VSI, i.e., $Y=0$.
In fig. 1, we also show $K_{vs}$ as a function of $M^*$, in the cases of $y=1$
and $y=5$, where $y=g_v^2\rho_0Y/m_v^2$ [12].
In those figures, it is seen that the VSI makes $K_{vs}$ smaller in the small
$M^*$ region.
$K_{vs}$ is almost independent on $y$ in the large $M^*$ region, where the
value of $M^*$ dominates the features of the equations of state.
As a result, the uncertainty of $K_{vs}$ in changing $M^*$ becomes smaller.
In the case of $K=300$MeV and $K_c=-3.990$MeV, $K_{vs}=-298\sim -282(-303\sim
-289)$MeV with $y=1.0(5.0)$.
In the case of $K=250$MeV and $K_c=-0.7065$MeV, $K_{vs}=50\sim 147(49\sim
86)$MeV with $y=1.0(5.0)$.
In each case, $K_{vs}$ with $y=5.0$ is not much different from that with
$y=1.0$.
$K_{vs}$ is hardly changed, if $y$ increases much more.
In table II(b), we summarize the range of the calculated $K_{vs}$ in the cases
of $y=5.0$.
Comparing this table with table I, we see that $K=300\pm 50$MeV is necessary to
account for the empirical values of $K$, $K_c$, and $K_{vs}$ at the same time,
as is in the case of no VSI.
Although the VSI makes $K_{vs}$ smaller (more negative), the change of $K_{vs}$
is comparable to the magnitude of the empirical error bars at the empirical
available value of $K(\sim 300$MeV), and, therefore, the conclusion is hardly
changed.
The result is also independent on the detail of $U(\phi )$ as in the case of no
VSI.

In summary, we have studied $K$, $K_c$, and $K_{vs}$ by using the relativistic
mean field theories with the nonlinear $\sigma$ term and with the nonlinear
$\sigma$ and $\omega$ terms, with aid of the scaling model.
It is found that, in both cases, $K=300\pm 50$MeV is $necessary$ to account for
the empirical values of $K$, $K_c$, and $K_{vs}$ at the same time.
The result is independent on the detail descriptions of $U(\phi )$ and is
almost independent on the strength of VSI.
It seems that this conclusion is not drastically changed, if we use any type of
the relativistic mean-field theory and the scaling model, since the calculated
$K_{vs}$ is most sensitive to the ratio $K'/K$, which is adjusted to the
empirical values.

\bigskip

\noindent
$Acknowledgment$: The authors gratefully acknowledge the computing time granted
by the Research Center for Nuclear Physics (RCNP).


\vfill\eject

\centerline{{\bf{References}}}

\bigskip

\noindent
[1] J.P. Blaizot, Phys. Rep. {\bf{64}} ,171(1980).

\noindent
[2] J.M. Pearson, Phys. Lett. B{\bf{271}} ,12(1991).

\noindent
[3] S. Shlomo and D.H. Youngblood, Phys. Rev. {\bf{C47}} ,529(1993).

\noindent
[4] S. Rudaz, P.J. Ellis, E.K. Heide and M. Prakash, Phys. Lett. {\bf{B285}},
183(1992).

\noindent
[5] E.K. Heide and S. Rudaz, Phys. Lett. {\bf{B262}}, 375(1991).

\noindent
[6] D. Von-Eiff, J.M. Pearson, W. Stocker and M.K. Weigel,

\noindent
Phys Lett. {\bf{B324}}, 279(1994).

\noindent
[7] D. Von-Eiff, J.M. Pearson, W. Stocker and M.K. Weigel,

\noindent
Phys. Rev. {\bf{C50}}, 831(1994).

\noindent
[8] D. Von-Eiff, W. Stocker and M.K. Weigel, Phys. Rev. {\bf{C50}}, 1436(1994).

\noindent
[9] A.R. Bodmer and C.E. Price, Nucl. Phys. {\bf{A505}}, 123(1989).

\noindent
[10] M.V. Stoitsov, P. Ring and M.M. Sharma, Phys. Rev. {\bf{C50}}, 1445(1994).

\noindent
[11] H. Kouno, N. Kakuta, N. Noda, K. Koide, T. Mitsumori, A. Hasegawa and M.
Nakano, Phys. Rev. {\bf{C51}}, 1754(1995).

\noindent
[12] H. Kouno, K. Koide, T. Mitsumori, N. Noda, A. Hasegawa and M. Nakano, to
be published in Phys. Rev. C.

\noindent
[13] J. Boguta and A.R. Bodmer, Nucl. Phys. {\bf{A292}}, 413(1977)

\noindent
[14] A.R. Bodmer, Nucl. Phys. {\bf{A526}}, 703(1991).

\noindent
[15] B.D. Serot, Phys. Lett. B{\bf{86}}, 146(1979):
%
B.D. Serot and J.D. Walecka, $The$ $Relativistic$ $Nuclear$ $Many$-$Body$
$Problem$ in: Advances in nuclear physics, vol. 16 (Plenum Press, New York,
1986).

\vfill\eject


\centerline{{\bf{Table and Figure Captions}}}

\bigskip

\noindent
Table I

\noindent
The sets of the empirical values of $K$, $K_c$ and $K_{vs}$ in the table 3 in
ref. [2]. (According to the conclusion in ref. [2], we only show the data in
the cases of $K=150\sim 350$MeV.) All quantities in the table are shown in MeV.

\bigskip

\noindent
Table II

\noindent
Range of the calculated $K_{vs}$ using the sets of $K$ and $K_c$ in table I as
inputs.
(a) The result in the the mean-field theory only with $\sigma$ meson
self-interaction.
(b) The result in the mean-field theory with $\sigma$ and $\omega$ mesons
self-interactions with $y=5.0$. :
In each table, "upper bound", "mean value", and "lower bound" mean that the
results are obtained by using the upper bound, the mean value, and the lower
bound of $K_c$ in table I, respectively.
All quantities in the table are shown in MeV.

\bigskip

\noindent
Fig. 1~~~~~$K_{vs}$ as a function of $M^*$.
(a) The cases of $K=300$MeV and $K_c=-3.990$MeV.
(b) The cases of $K=250$MeV and $K_c=-0.7065$MeV:
In each figure, the solid line is the result in the the mean-field theory only
with $\sigma$ meson self-interaction, and the dotted and the dashed lines are
the results in the mean-field theory with $\sigma$ and $\omega$ mesons
self-interactions with $y=1.0$ and $y=5.0$, respectively.

\vfill\eject


\large

\hspace*{-2cm}
  \begin{tabular}{cccccc}
                                    \hline
    \     & Set 1 & Set 2 & Set 3 & Set 4 & Set 5  \\ \hline
    \   $K$ & 150.0 & 200.0  & 250.0 & 300.0 & 350.0 \\
    \  $K_c$ & $5.861\pm2.06 $ & $2.577\pm2.06 $ & $-0.7065\pm2.06 $
     & $-3.990\pm2.06 $ &$-7.274\pm2.06 $ \\
    \ $K_{vs}$ & $66.83\pm101$ & $-46.94\pm101$ & $-160.7\pm101$
     & $-274.5\pm101$ & $-388.3\pm101$ \\ \hline
  \end{tabular}

\bigskip

\begin{center}
Table I
\end{center}

\bigskip

\bigskip

\bigskip


  \begin{tabular}{cccc}
                                    \hline
    \         & upper bound    & mean value     & lower bound    \\ \hline
    \ $K=150$ &  940$\sim$1738 &  728$\sim$1426 &  517$\sim$1113 \\
    \ $K=200$ &  601$\sim$1230 &  389$\sim$ 918 &  178$\sim$ 605 \\
    \ $K=250$ &  262$\sim$ 722 &   50$\sim$ 410 & -161$\sim$  97 \\
    \ $K=300$ &  -77$\sim$ 214 & -289$\sim$ -98 & -506$\sim$-411 \\
    \ $K=350$ & -418$\sim$-294 & -651$\sim$-606 & -921$\sim$-840 \\ \hline
  \end{tabular}

\bigskip

\begin{center}
      Table II(a)
\end{center}

\bigskip

\bigskip

\bigskip


  \begin{tabular}{cccc}
                                    \hline
    \         &  upper bound   & mean value     & lower bound    \\ \hline
    \ $K=150$ &  939$\sim$1103 &  728$\sim$ 863 &  516$\sim$ 623 \\
    \ $K=200$ &  600$\sim$ 714 &  389$\sim$ 474 &  177$\sim$ 235 \\
    \ $K=250$ &  261$\sim$ 325 &   49$\sim$  86 & -165$\sim$-154 \\
    \ $K=300$ &  -80$\sim$ -63 & -303$\sim$-289 & -542$\sim$-500 \\
    \ $K=350$ & -451$\sim$-416 & -691$\sim$-628 & -931$\sim$-839 \\ \hline
  \end{tabular}

\bigskip

\begin{center}
      Table II(b)
\end{center}

\end{document}